\documentclass[nofootinbib,prd]{revtex4}%
\usepackage{amsmath}
\usepackage{amsfonts}
\usepackage{amssymb}
\usepackage{graphicx}%
\begin{document}
\title{Modified Dispersion Relations lead to a finite Zero Point Gravitational Energy}
\author{Remo Garattini}
\email{Remo.Garattini@unibg.it}
\affiliation{Universit\`{a} degli Studi di Bergamo, Facolt\`{a} di Ingegneria,}
\affiliation{Viale Marconi 5, 24044 Dalmine (Bergamo) Italy}
\affiliation{and I.N.F.N. - sezione di Milano, Milan, Italy.}
\author{Gianluca Mandanici}
\email{Gianluca.Mandanici@unibg.it}
\affiliation{Universit\`{a} degli Studi di Bergamo, Facolt\`{a} di Ingegneria, Viale
Marconi 5, 24044 Dalmine (Bergamo) ITALY.}

\begin{abstract}
We compute the Zero Point Energy in a spherically symmetric background
distorted at high energy as predicted by \textit{Gravity's Rainbow}. In this
context we setup a Sturm-Liouville problem with the cosmological constant
considered as the associated eigenvalue. The eigenvalue equation is a
reformulation of the Wheeler-DeWitt equation. With the help of a canonical
decomposition, we find that the relevant contribution to one loop is given by
the graviton quantum fluctuations around the given background. By means of a
variational approach based on gaussian trial functionals, we find that the
ordinary divergences can here be handled by an appropriate choice of the
rainbow's functions, in contrast to what happens in other conventional
approaches. A final discussion on the connection of our result with the
observed cosmological constant is also reported.

\end{abstract}
\maketitle

\section{Introduction}

The idea of promoting General Relativity to a quantum level, together with a
solution of the Cosmological Constant Problem\textbf{,} is one of the big
challenges of our century. Indeed a satisfying \textit{Quantum Gravity theory}
does not exist yet and the enormous gap of $10^{120}$ orders of magnitude
between the predicted theoretical value of the cosmological constant and the
observed one has not yet found a compelling explanation. If the quantum
description of Nature is appropriate for every force, it should be applicable
even to the gravitational force described by General Relativity. But perhaps,
General Relativity as it stands, requires a change In this respect, various
proposals on how the fundamental aspects of special relativity can be modified
at very high energies have been done. Among these proposals, particularly
promising appears to be the one known as \textit{Doubly Special Relativity}
(DSR)\cite{GAC}\textbf{.} One of the characterizing DSR effects is that the
usual dispersion relation of a massive particle of mass $m$ is modified into
the following expression%
\begin{equation}
E^{2}g_{1}^{2}\left(  E/E_{P}\right)  -p^{2}g_{2}^{2}\left(  E/E_{P}\right)
=m^{2}, \label{mdisp}%
\end{equation}
where $g_{1}\left(  E/E_{P}\right)  $ and $g_{2}\left(  E/E_{P}\right)  $ are
two arbitrary functions which have the following property%
\begin{equation}
\lim_{E/E_{P}\rightarrow0}g_{1}\left(  E/E_{P}\right)  =1\qquad\text{and}%
\qquad\lim_{E/E_{P}\rightarrow0}g_{2}\left(  E/E_{P}\right)  =1. \label{lim}%
\end{equation}
Thus, the usual dispersion relation is recovered at low energies. Of course
the first ideas of DSR were minted for flat space. However, nothing forbids to
consider a curved background and therefore to enter into the realm of General
Relativity. From this point of view, Magueijo and Smolin\cite{MagSmo} proposed
that the energy-momentum tensor and the Einstein's Field Equations were
modified with the introduction of a one parameter family of equations%
\begin{equation}
G_{\mu\nu}\left(  E\right)  =8\pi G\left(  E\right)  T_{\mu\nu}\left(
E\right)  +g_{\mu\nu}\Lambda\left(  E\right)  ,
\end{equation}
where $G\left(  E\right)  $ is an energy dependent Newton's constant, defined
so that $G\left(  0\right)  $ is the low-energy Newton's constant. Similarly
we have an energy dependent cosmological constant $\Lambda\left(  E\right)  $
leading to the \textit{rainbow} version of the Schwarzschild line element%
\begin{equation}
ds^{2}=-\left(  1-\frac{2MG\left(  0\right)  }{r}\right)  \frac{d\tilde{t}%
^{2}}{g_{1}^{2}\left(  E/E_{P}\right)  }+\frac{d\tilde{r}^{2}}{\left(
1-\frac{2MG\left(  0\right)  }{r}\right)  g_{2}^{2}\left(  E/E_{P}\right)
}+\frac{\tilde{r}^{2}}{g_{2}^{2}\left(  E/E_{P}\right)  }\left(  d\theta
^{2}+\sin^{2}\theta d\phi^{2}\right)  . \label{line}%
\end{equation}
Since the functions $g_{1}\left(  E/E_{P}\right)  $ and $g_{2}\left(
E/E_{P}\right)  $ come into play when the energy $E$ is comparable with
$E_{P}$, it is likely that they modify the UV behavior in the same way as
Generalized Uncertainty Principle and Noncommutative Geometry (NCG) do,
respectively. If the effect of Generalized Uncertainty Principle and NCG is to
modify the Liouville measure $d^{3}xd^{3}k$ from one side,\textbf{ }and to
introduce a granularity from the other side, the \textit{rainbow metric
}$\left(  \ref{line}\right)  $ should be able to introduce a natural UV
regulator hidden into the arbitrary\textit{ }functions $g_{1}\left(
E/E_{P}\right)  $ and $g_{2}\left(  E/E_{P}\right)  $. An encouraging partial
answer has been obtained in Ref.\cite{RemoPLB}, where an application of
Rainbow's Gravity to black hole entropy computation has been considered. In
that paper the UV regulator, namely the \textit{brick wall,} has been
eliminated with the help of the following choice of $g_{1}\left(
E/E_{P}\right)  $ and $g_{2}\left(  E/E_{P}\right)  $%
\begin{equation}
\frac{g_{1}\left(  E/E_{P}\right)  }{g_{2}\left(  E/E_{P}\right)  }%
=\exp\left(  -\frac{E}{E_{P}}\right)  . \label{rap}%
\end{equation}
An interesting test to see \textit{Rainbow's Gravity} at work again, should be
the computation of Zero Point Energy (ZPE). Nevertheless, we have to remark
that any computation of ZPE leads to a regularization and subsequently to a
renormalization process in order to have finite physical quantities. Therefore
the purpose of the paper is to show that with the introduction of appropriate
Rainbow's Gravity functions it is possible to overpass the renormalization
problem. Of course this proposal does not represent a complete cure to have a
finite theory of Quantum Gravity but rather it suggests how the modification
of some basic principles like the introduction of an energy dependent metric
can lead to unexpected results such as the avoidance of a renormalization
scheme. In ordinary gravity the computation of ZPE for quantum fluctuations of
the \textit{pure gravitational field} can be extracted by rewriting the
Wheeler-DeWitt equation (WDW)\cite{DeWitt} in a form which looks like an
expectation value computation\cite{Remo}. We remind the reader that the WDW
equation is the quantum version of the classical constraint which guarantees
the invariance under time reparametrization. Its original form with the
cosmological term included is described by%
\begin{equation}
\mathcal{H}\Psi=\left[  \left(  2\kappa\right)  G_{ijkl}\pi^{ij}\pi^{kl}%
-\frac{\sqrt{g}}{2\kappa}\!{}\!\left(  \,\!^{3}R-2\Lambda\right)  \right]
\Psi=0. \label{WDW}%
\end{equation}
If we multiply Eq.$\left(  \ref{WDW}\right)  $ by $\Psi^{\ast}\left[
g_{ij}\right]  $ and functionally integrate over the three spatial metric
$g_{ij}$, we can write\footnote{See also Ref.\cite{CG} for an application of
the method to a $f\left(  R\right)  $ theory.}\cite{Remo}%
\begin{equation}
\frac{1}{V}\frac{\int\mathcal{D}\left[  g_{ij}\right]  \Psi^{\ast}\left[
g_{ij}\right]  \int_{\Sigma}d^{3}x\hat{\Lambda}_{\Sigma}\Psi\left[
g_{ij}\right]  }{\int\mathcal{D}\left[  g_{ij}\right]  \Psi^{\ast}\left[
g_{ij}\right]  \Psi\left[  g_{ij}\right]  }=\frac{1}{V}\frac{\left\langle
\Psi\left\vert \int_{\Sigma}d^{3}x\hat{\Lambda}_{\Sigma}\right\vert
\Psi\right\rangle }{\left\langle \Psi|\Psi\right\rangle }=-\frac{\Lambda
}{\kappa}, \label{VEV}%
\end{equation}
where we have also integrated over the hypersurface $\Sigma$ and we have
defined%
\begin{equation}
V=\int_{\Sigma}d^{3}x\sqrt{g}%
\end{equation}
as the volume of the hypersurface $\Sigma$ with%
\begin{equation}
\hat{\Lambda}_{\Sigma}=\left(  2\kappa\right)  G_{ijkl}\pi^{ij}\pi^{kl}%
-\sqrt{g}^{3}R/\left(  2\kappa\right)  . \label{LambdaSigma}%
\end{equation}
In this form, Eq.$\left(  \ref{VEV}\right)  $ can be used to compute ZPE
provided that $\Lambda/\kappa$ be considered as an eigenvalue of $\hat
{\Lambda}_{\Sigma}$, namely the WDW equation is transformed into an
expectation value computation. In Eq.$\left(  \ref{WDW}\right)  $, $G_{ijkl}$
is the super-metric, $\pi^{ij}$ is the super-momentum,$^{3}R$ is the scalar
curvature in three dimensions and $\Lambda$ is the cosmological constant,
while $\kappa=8\pi G$ with $G$ the Newton's constant. Nevertheless, solving
Eq.$\left(  \ref{VEV}\right)  $ is a quite impossible task, therefore we are
oriented to use a variational approach with trial wave functionals. The
related boundary conditions are dictated by the choice of the trial wave
functionals which, in our case\textbf{,} are of the Gaussian type. Different
types of wave functionals correspond to different boundary conditions. The
choice of a Gaussian wave functional is justified by the fact that ZPE should
be described by a good candidate of the \textquotedblleft\textit{vacuum
state}\textquotedblright. To fix the ideas, a variant of the line element
$\left(  \ref{line}\right)  $ will be considered%
\begin{equation}
ds^{2}=-N^{2}\left(  r\right)  \frac{dt^{2}}{g_{1}^{2}\left(  E\right)
}+\frac{dr^{2}}{\left(  1-\frac{b\left(  r\right)  }{r}\right)  g_{2}%
^{2}\left(  E\right)  }+\frac{r^{2}}{g_{2}^{2}\left(  E\right)  }\left(
d\theta^{2}+\sin^{2}\theta d\phi^{2}\right)  , \label{dS}%
\end{equation}
where $N$ is the lapse function and $b\left(  r\right)  $ is subject to the
only condition $b\left(  r_{t}\right)  =r_{t}$. Metric $\left(  \ref{dS}%
\right)  $, will be our cornerstone of the whole paper which is organized as
follows. In section \ref{p2}, we derive the Hamiltonian constraint in presence
of the background $\left(  \ref{dS}\right)  $, in section \ref{p3} we compute
the ZPE of quantum fluctuations around the background $\left(  \ref{dS}%
\right)  $ and with the help of an appropriate choice of the functions
$g_{1}\left(  E/E_{P}\right)  $ and $g_{2}\left(  E/E_{P}\right)  $\textbf{,}
we will show that the UV divergences of ZPE disappear. We summarize and
conclude in section \ref{p4}. Units in which $\hbar=c=k=1$ are used throughout
the paper.

\section{The Hamiltonian Constraint in Rainbow's Gravity}

\label{p2}In order to use Eq.$\left(  \ref{VEV}\right)  $ for the metric
$\left(  \ref{dS}\right)  $, we need to understand how the WDW modifies when
the functions $g_{1}\left(  E/E_{Pl}\right)  $ and $g_{2}\left(
E/E_{Pl}\right)  $ distort the background. It is therefore necessary to
understand how some basic ingredients change under the transformation of the
line element $\left(  \ref{dS}\right)  $. The form of the background is such
that the \textit{shift function}
\begin{equation}
N^{i}=-Nu^{i}=g_{0}^{4i}=0
\end{equation}
vanishes, while $N$ is the previously defined \textit{lapse function}. Thus
the definition of $K_{ij}$ implies
\begin{equation}
K_{ij}=-\frac{\dot{g}_{ij}}{2N}=\frac{g_{1}\left(  E\right)  }{g_{2}%
^{2}\left(  E\right)  }\tilde{K}_{ij},
\end{equation}
where the dot denotes differentiation with respect to the time $t$ and the
tilde indicates the quantity computed in absence of rainbow's functions
$g_{1}\left(  E\right)  $ and $g_{2}\left(  E\right)  $. For simplicity, we
have set $E_{P}=1$ in $g_{1}\left(  E/E_{P}\right)  $ and $g_{2}\left(
E/E_{P}\right)  $ throughout the paragraph. The trace of the extrinsic
curvature, therefore becomes%
\begin{equation}
K=g^{ij}K_{ij}=g_{1}\left(  E\right)  \tilde{K}%
\end{equation}
and the momentum $\pi^{ij}$ conjugate to the three-metric $g_{ij}$ of $\Sigma$
is%
\begin{equation}
\pi^{ij}=\frac{\sqrt{g}}{2\kappa}\left(  Kg^{ij}-K^{ij}\right)  =\frac
{g_{1}\left(  E\right)  }{g_{2}\left(  E\right)  }\tilde{\pi}^{ij}.
\end{equation}
Now, we have enough information to define the WDW equation for a background
described by $\left(  \ref{dS}\right)  $. From Eq.$\left(  \ref{WDW}\right)  $
we find that $\mathcal{H}\Psi=0$ becomes%
\begin{equation}
\mathcal{H}\Psi=\left[  \left(  2\kappa\right)  \frac{g_{1}^{2}\left(
E\right)  }{g_{2}^{3}\left(  E\right)  }\tilde{G}_{ijkl}\tilde{\pi}^{ij}%
\tilde{\pi}^{kl}\mathcal{-}\frac{\sqrt{\tilde{g}}}{2\kappa g_{2}\left(
E\right)  }\!{}\!\left(  \tilde{R}-\frac{2\Lambda_{c}}{g_{2}^{2}\left(
E\right)  }\right)  \right]  \Psi=0, \label{AccaR}%
\end{equation}
where we have used the following property on $R$%
\begin{equation}
R=g^{ij}R_{ij}=g_{2}^{2}\left(  E\right)  \tilde{R} \label{RR}%
\end{equation}
and where
\begin{equation}
G_{ijkl}=\frac{1}{2\sqrt{g}}\left(  g_{ik}g_{jl}+g_{il}g_{jk}-g_{ij}%
g_{kl}\right)  =\frac{\tilde{G}_{ijkl}}{g_{2}\left(  E\right)  }.
\end{equation}
Therefore, in presence of Rainbow's Gravity, we find that Eq.$\left(
\ref{VEV}\right)  $ becomes%
\begin{equation}
\frac{g_{2}^{3}\left(  E\right)  }{\tilde{V}}\frac{\left\langle \Psi\left\vert
\int_{\Sigma}d^{3}x\tilde{\Lambda}_{\Sigma}\right\vert \Psi\right\rangle
}{\left\langle \Psi|\Psi\right\rangle }=-\frac{\Lambda_{c}}{\kappa},
\label{WDW3}%
\end{equation}
where%
\begin{equation}
\tilde{\Lambda}_{\Sigma}=\left(  2\kappa\right)  \frac{g_{1}^{2}\left(
E\right)  }{g_{2}^{3}\left(  E\right)  }\tilde{G}_{ijkl}\tilde{\pi}^{ij}%
\tilde{\pi}^{kl}\mathcal{-}\frac{\sqrt{\tilde{g}}\tilde{R}}{\left(
2\kappa\right)  g_{2}\left(  E\right)  }\!{}\!. \label{LambdaR}%
\end{equation}

We can gain more information if we consider $g_{ij}=\bar{g}_{ij}+h_{ij}$,
where $\bar{g}_{ij}$ is the background metric and $h_{ij}$ is a quantum
fluctuation around the background. Thus Eq.$\left(  \ref{WDW3}\right)  $ can
be expanded in terms of $h_{ij}$. Since the kinetic part of $\hat{\Lambda
}_{\Sigma}$ is quadratic in the momenta, we only need to expand the
three-scalar curvature $\int d^{3}x\sqrt{g}{}^{3}R$ up to the quadratic order.
However, to proceed with the computation, we also need an orthogonal
decomposition on the tangent space of 3-metric
deformations\cite{Vassilevich,Quad}:%

\begin{equation}
h_{ij}=\frac{1}{3}\left(  \sigma+2\nabla\cdot\xi\right)  g_{ij}+\left(
L\xi\right)  _{ij}+h_{ij}^{\bot}.\label{p21a}%
\end{equation}
The operator $L$ maps $\xi_{i}$ into symmetric tracefree tensors
\begin{equation}
\left(  L\xi\right)  _{ij}=\nabla_{i}\xi_{j}+\nabla_{j}\xi_{i}-\frac{2}%
{3}g_{ij}\left(  \nabla\cdot\xi\right)  ,
\end{equation}
$h_{ij}^{\bot}$ is the traceless-transverse component of the perturbation
(TT), namely
\begin{equation}
g^{ij}h_{ij}^{\bot}=0,\qquad\nabla^{i}h_{ij}^{\bot}=0
\end{equation}
and $h$ is the trace of $h_{ij}$. It is immediate to recognize that the trace
element $\sigma=h-2\left(  \nabla\cdot\xi\right)  $ is gauge invariant. It is
straightforward to see that the gauge invariant decomposition $\left(
\ref{p21a}\right)  $ does not change, when we consider the rainbow's metric
$\left(  \ref{dS}\right)  $. Therefore, following the results of
Ref.\cite{Remo1}, we can use the final expression%
\begin{equation}
\frac{1}{V}\frac{\left\langle \Psi^{\bot}\left\vert \int_{\Sigma}d^{3}x\left[
\hat{\Lambda}_{\Sigma}^{\bot}\right]  ^{\left(  2\right)  }\right\vert
\Psi^{\bot}\right\rangle }{\left\langle \Psi^{\bot}|\Psi^{\bot}\right\rangle
}+\frac{1}{V}\frac{\left\langle \Psi^{\sigma}\left\vert \int_{\Sigma}%
d^{3}x\left[  \hat{\Lambda}_{\Sigma}^{\sigma}\right]  ^{\left(  2\right)
}\right\vert \Psi^{\sigma}\right\rangle }{\left\langle \Psi^{\sigma}%
|\Psi^{\sigma}\right\rangle }=-\frac{\Lambda_{c}}{\kappa}.\label{lambda0_2a}%
\end{equation}
Note that in the expansion of $\int_{\Sigma}d^{3}x\sqrt{g}{}R$ to second order
in terms of $h_{ij}$, a coupling term between the TT component and the scalar
one remains. However, the Gaussian integration does not allow such a mixing
which has to be introduced with an appropriate wave functional. Extracting the
TT tensor contribution from Eq.$\left(  \ref{WDW3}\right)  $, we find%
\begin{equation}
\hat{\Lambda}_{\Sigma}^{\bot}=\frac{g_{2}^{3}\left(  E\right)  }{4\tilde{V}%
}\int_{\Sigma}d^{3}x\sqrt{\overset{\sim}{\bar{g}}}\tilde{G}^{ijkl}\left[
\left(  2\kappa\right)  \frac{g_{1}^{2}\left(  E\right)  }{g_{2}^{3}\left(
E\right)  }\tilde{K}^{-1\bot}\left(  x,x\right)  _{ijkl}+\frac{1}{\left(
2\kappa\right)  g_{2}\left(  E\right)  }\!{}\left(  \tilde{\bigtriangleup
}_{L\!}^{m}\tilde{K}^{\bot}\left(  x,x\right)  \right)  _{ijkl}\right]
.\label{p22}%
\end{equation}
The origin of the operator $\tilde{\bigtriangleup}_{L\!}^{m}$ comes from%
\begin{equation}
\left(  \hat{\bigtriangleup}_{L\!}^{m}\!{}h^{\bot}\right)  _{ij}=\left(
\bigtriangleup_{L\!}\!{}h^{\bot}\right)  _{ij}-4R{}_{i}^{k}\!{}h_{kj}^{\bot
}+\text{ }^{3}R{}\!{}h_{ij}^{\bot},\label{M Lichn}%
\end{equation}
which is the modified Lichnerowicz operator where $\bigtriangleup_{L}$is the
Lichnerowicz operator defined by%
\begin{equation}
\left(  \bigtriangleup_{L}h\right)  _{ij}=\bigtriangleup h_{ij}-2R_{ikjl}%
h^{kl}+R_{ik}h_{j}^{k}+R_{jk}h_{i}^{k}\qquad\bigtriangleup=-\nabla^{a}%
\nabla_{a}.\label{DeltaL}%
\end{equation}
$G^{ijkl}$ represents the inverse DeWitt metric without the $\sqrt{g}$ factor
and all indices run from one to three. Note that the term%
\begin{equation}
-4R{}_{i}^{k}\!{}h_{kj}^{\bot}+^{3}R{}\!{}h_{ij}^{\bot}%
\end{equation}
disappears in four dimensions when we use a background which is a solution of
the Einstein's field equations without matter contribution. The
\textquotedblleft$\sim$\textquotedblright\ symbol in Eq.$\left(
\ref{p22}\right)  $ means that we have rescaled every piece in Eq.$\left(
\ref{WDW3}\right)  $, evaluated at second order. Moreover, although the
expression of $\tilde{\Lambda}_{\Sigma}$ explicitly shows how globally changes
the operator $\hat{\Lambda}_{\Sigma}$, when we consider the following
eigenvalue equation%
\begin{equation}
\left(  \hat{\bigtriangleup}_{L\!}^{m}\!{}h^{\bot}\right)  _{ij}=E^{2}%
h_{ij}^{\bot},\label{EE}%
\end{equation}
we find that
\begin{equation}
\left(  \tilde{\bigtriangleup}_{L\!}^{m}\!{}\tilde{h}^{\bot}\right)  _{ij}%
\!{}=\frac{E^{2}}{g_{2}^{2}\left(  E\right)  }\tilde{h}_{ij}^{\bot},
\end{equation}
in order to reestablish the correct way of transformation of the perturbation.
Then, the propagator $K^{\bot}\left(  x,x\right)  _{iakl}$ can be represented
as
\begin{equation}
K^{\bot}\left(  \overrightarrow{x},\overrightarrow{y}\right)  _{iakl}%
=\tilde{K}^{\bot}\left(  \overrightarrow{x},\overrightarrow{y}\right)
_{iakl}=\sum_{\tau}\frac{\tilde{h}_{ia}^{\left(  \tau\right)  \bot}\left(
\overrightarrow{x}\right)  \tilde{h}_{kl}^{\left(  \tau\right)  \bot}\left(
\overrightarrow{y}\right)  }{2\lambda\left(  \tau\right)  g_{2}^{4}\left(
E\right)  },\label{proptt}%
\end{equation}
where $\tilde{h}_{ia}^{\left(  \tau\right)  \bot}\left(  \overrightarrow
{x}\right)  $ are the eigenfunctions of $\tilde{\bigtriangleup}_{L\!}^{m}$.
$\tau$ denotes a complete set of indices and $\lambda\left(  \tau\right)  $
are a set of variational parameters to be determined by the minimization of
Eq.$\left(  \ref{p22}\right)  $. The expectation value of $\hat{\Lambda
}_{\Sigma}^{\bot}$ is easily obtained by inserting the form of the propagator
into Eq.$\left(  \ref{p22}\right)  $ and minimizing with respect to the
variational function $\lambda\left(  \tau\right)  $. Thus the total one loop
energy density for TT tensors becomes%
\begin{equation}
\frac{\Lambda}{8\pi G}=-\frac{1}{2}\sum_{\tau}g_{1}\left(  E\right)
g_{2}\left(  E\right)  \left[  \sqrt{E_{1}^{2}\left(  \tau\right)  }%
+\sqrt{E_{2}^{2}\left(  \tau\right)  }\right]  .\label{1loop}%
\end{equation}
The above expression makes sense only for $E_{i}^{2}\left(  \tau\right)  >0$,
where $E_{i}$ are the eigenvalues of $\tilde{\bigtriangleup}_{L\!}^{m}$. With
the help of Regge and Wheeler representation\cite{Regge Wheeler}, the
eigenvalue equation $\left(  \ref{EE}\right)  $ can be reduced to%
\begin{equation}
\left[  -\frac{d^{2}}{dx^{2}}+\frac{l\left(  l+1\right)  }{r^{2}}+m_{i}%
^{2}\left(  r\right)  \right]  f_{i}\left(  x\right)  =\frac{E_{i,l}^{2}%
}{g_{2}^{2}\left(  E\right)  }f_{i}\left(  x\right)  \quad i=1,2\quad
,\label{p34}%
\end{equation}
where we have used reduced fields of the form $f_{i}\left(  x\right)
=F_{i}\left(  x\right)  /r$ and where we have defined two r-dependent
effective masses $m_{1}^{2}\left(  r\right)  $ and $m_{2}^{2}\left(  r\right)
$%
\begin{equation}
\left\{
\begin{array}
[c]{c}%
m_{1}^{2}\left(  r\right)  =\frac{6}{r^{2}}\left(  1-\frac{b\left(  r\right)
}{r}\right)  +\frac{3}{2r^{2}}b^{\prime}\left(  r\right)  -\frac{3}{2r^{3}%
}b\left(  r\right)  \\
\\
m_{2}^{2}\left(  r\right)  =\frac{6}{r^{2}}\left(  1-\frac{b\left(  r\right)
}{r}\right)  +\frac{1}{2r^{2}}b^{\prime}\left(  r\right)  +\frac{3}{2r^{3}%
}b\left(  r\right)
\end{array}
\right.  \quad\left(  r\equiv r\left(  x\right)  \right)  .\label{masses}%
\end{equation}
In order to use the WKB approximation, from Eq.$\left(  \ref{p34}\right)  $ we
can extract two r-dependent radial wave numbers%
\begin{equation}
k_{i}^{2}\left(  r,l,\omega_{i,nl}\right)  =\frac{E_{i,nl}^{2}}{g_{2}%
^{2}\left(  E\right)  }-\frac{l\left(  l+1\right)  }{r^{2}}-m_{i}^{2}\left(
r\right)  \quad i=1,2\quad.\label{kTT}%
\end{equation}

\section{One loop energy in an ordinary spherically symmetric background}

\label{p3}It is now possible to explicitly evaluate Eq.$\left(  \ref{1loop}%
\right)  $ in terms of the effective mass. To further proceed we use the
W.K.B. method used by `t Hooft in the brick wall problem\cite{tHooft} and we
count the number of modes with frequency less than $\omega_{i}$, $i=1,2$. This
is given approximately by%
\begin{equation}
\tilde{g}\left(  E_{i}\right)  =\int_{0}^{l_{\max}}\nu_{i}\left(
l,E_{i}\right)  \left(  2l+1\right)  dl, \label{p41}%
\end{equation}
where $\nu_{i}\left(  l,E_{i}\right)  $, $i=1,2$ is the number of nodes in the
mode with $\left(  l,E_{i}\right)  $, such that $\left(  r\equiv r\left(
x\right)  \right)  $
\begin{equation}
\nu_{i}\left(  l,E_{i}\right)  =\frac{1}{\pi}\int_{-\infty}^{+\infty}%
dx\sqrt{k_{i}^{2}\left(  r,l,E_{i}\right)  }. \label{p42}%
\end{equation}
Here it is understood that the integration with respect to $x$ and $l_{\max}$
is taken over those values which satisfy $k_{i}^{2}\left(  r,l,E_{i}\right)
\geq0,$ $i=1,2$. With the help of Eqs.$\left(  \ref{p41},\ref{p42}\right)  $,
Eq.$\left(  \ref{1loop}\right)  $ leads to%
\begin{equation}
\frac{\Lambda}{8\pi G}=-\frac{1}{\pi}\sum_{i=1}^{2}\int_{0}^{+\infty}%
E_{i}g_{1}\left(  E\right)  g_{2}\left(  E\right)  \frac{d\tilde{g}\left(
E_{i}\right)  }{dE_{i}}dE_{i}. \label{tot1loop}%
\end{equation}
This is the graviton contribution to the induced cosmological constant to one
loop. The explicit evaluation of the density of states yields%
\[
\frac{d\tilde{g}(E_{i})}{dE_{i}}=\int\frac{\partial\nu(l{,}E_{i})}{\partial
E_{i}}(2l+1)dl=\frac{1}{\pi}\int_{-\infty}^{+\infty}dx\int_{0}^{l_{\max}}%
\frac{(2l+1)}{\sqrt{k^{2}(r,l,E)}}\frac{d}{dE_{i}}\left(  \frac{E_{i}^{2}%
}{g_{2}^{2}\left(  E\right)  }-m_{i}^{2}\left(  r\right)  \right)  dl
\]%
\begin{equation}
=\frac{2}{\pi}\int_{-\infty}^{+\infty}dxr^{2}\frac{d}{dE_{i}}\left(
\frac{E_{i}^{2}}{g_{2}^{2}\left(  E\right)  }-m_{i}^{2}\left(  r\right)
\right)  \sqrt{\frac{E_{i}^{2}}{g_{2}^{2}\left(  E\right)  }-m_{i}^{2}\left(
r\right)  }=\frac{4}{3\pi}\int_{-\infty}^{+\infty}dxr^{2}\frac{d}{dE_{i}%
}\left(  \frac{E_{i}^{2}}{g_{2}^{2}\left(  E\right)  }-m_{i}^{2}\left(
r\right)  \right)  ^{\frac{3}{2}}. \label{states}%
\end{equation}
Plugging expression $\left(  \ref{states}\right)  $ into Eq.$\left(
\ref{tot1loop}\right)  $ and dividing for a volume factor, we obtain%
\begin{equation}
\frac{\Lambda}{8\pi G}=-\frac{1}{3\pi^{2}}\sum_{i=1}^{2}\int_{E^{\ast}%
}^{+\infty}E_{i}g_{1}\left(  E\right)  g_{2}\left(  E\right)  \frac{d}{dE_{i}%
}\sqrt{\left(  \frac{E_{i}^{2}}{g_{2}^{2}\left(  E\right)  }-m_{i}^{2}\left(
r\right)  \right)  ^{3}}dE_{i}, \label{LoverG}%
\end{equation}
where $E^{\ast}$ is the value which annihilates the argument of the root. In
the previous equation, we have included an additional $4\pi$ factor coming
from the angular integration and we have assumed that the effective mass does
not depend on the energy $E$. To further proceed, we can see what happens to
the expression $\left(  \ref{LoverG}\right)  $ for some specific forms of
$g_{1}\left(  E/E_{P}\right)  $ and $g_{2}\left(  E/E_{P}\right)  $. One
popular choice is given by%
\begin{equation}
g_{1}\left(  E/E_{P}\right)  =1-\eta\left(  E/E_{P}\right)  ^{n}%
\qquad\text{and}\qquad g_{2}\left(  E/E_{P}\right)  =1,
\end{equation}
where $\eta$ is a dimensionless parameter and $n$ is an integer\cite{g1g2}.
Nevertheless, the above choice does not allow a finite result in Eq.$\left(
\ref{LoverG}\right)  $ and therefore will be discarded. Thus the choice of the
possible forms of $g_{1}\left(  E/E_{P}\right)  $ and $g_{2}\left(
E/E_{P}\right)  $ is strongly restricted by convergence criteria. We have
hitherto used a generic form of the background. We now fix the attention on
some backgrounds which have the following property%
\begin{equation}
m_{0}^{2}\left(  r\right)  =m_{2}^{2}\left(  r\right)  =-m_{1}^{2}\left(
r\right)  ,\qquad\forall r\in\left(  r_{t},r_{1}\right)  . \label{cond}%
\end{equation}

For example, the Schwarzschild background represented by the choice $b\left(
r\right)  =r_{t}=2MG$ satisfies the property $\left(  \ref{cond}\right)  $ in
the range $r\in\left[  r_{t},5r_{t}/2\right]  $. Similar backgrounds are the
Schwarzschild-de Sitter and Schwarzschild-Anti de Sitter. On the other hand
other backgrounds, like de Sitter, Anti-de Sitter and Minkowski have the
property%
\begin{equation}
m_{0}^{2}\left(  r\right)  =m_{2}^{2}\left(  r\right)  =m_{1}^{2}\left(
r\right)  ,\qquad\forall r\in\left(  r_{t},\infty\right)  . \label{equal}%
\end{equation}

In case condition $\left(  \ref{cond}\right)  $ holds, Eq.$\left(
\ref{LoverG}\right)  $ becomes%
\begin{equation}
\frac{\Lambda}{8\pi G}=-\frac{1}{3\pi^{2}}\left(  I_{+}+I_{-}\right)  ,
\label{LoveG}%
\end{equation}
where%
\begin{equation}
I_{+}=\int_{0}^{\infty}\left(  Eg_{1}\left(  E/E_{P}\right)  g_{2}\left(
E/E_{P}\right)  \right)  \frac{d}{dE}\left(  \frac{E^{2}}{g_{2}^{2}\left(
E/E_{P}\right)  }+m_{0}^{2}\left(  r\right)  \right)  ^{\frac{3}{2}}dE
\label{I+}%
\end{equation}
and%
\begin{equation}
I_{-}=\int_{E^{\ast}}^{\infty}\left(  Eg_{1}\left(  E/E_{P}\right)
g_{2}\left(  E/E_{P}\right)  \right)  \frac{d}{dE}\left(  \frac{E^{2}}%
{g_{2}^{2}\left(  E/E_{P}\right)  }-m_{0}^{2}\left(  r\right)  \right)
^{\frac{3}{2}}dE. \label{I-}%
\end{equation}
Instead, in case condition $\left(  \ref{equal}\right)  $ holds, Eq.$\left(
\ref{LoverG}\right)  $ becomes%
\begin{equation}
\frac{\Lambda}{8\pi G}=-\frac{2}{3\pi^{2}}I_{-}. \label{LoveGeq}%
\end{equation}
We begin to look at Eq.$\left(  \ref{LoveG}\right)  $. It is immediate to see
that integrals $I_{+}$ and $I_{-}$ can be easily solved for a very particular
choice. Indeed, if we set%
\begin{equation}
g_{2}^{-2}(E/E_{P})=g_{1}(E/E_{P}) \label{symm}%
\end{equation}
we find that $I_{+}$ and $I_{-}$ take the form:%
\begin{equation}
I_{+}=3\int_{0}^{\infty}\left(  \frac{E}{g_{2}\left(  E/E_{P}\right)
}\right)  ^{2}\frac{d}{dE}\left(  \frac{E}{g_{2}\left(  E/E_{P}\right)
}\right)  \sqrt{\left(  \frac{E}{g_{2}\left(  E/E_{P}\right)  }\right)
^{2}+m_{0}^{2}\left(  r\right)  }dE \label{I++}%
\end{equation}
and%
\begin{equation}
I_{-}=3\int_{E^{\ast}}^{\infty}\left(  \frac{E}{g_{2}\left(  E/E_{P}\right)
}\right)  ^{2}\frac{d}{dE}\left(  \frac{E}{g_{2}\left(  E/E_{P}\right)
}\right)  \sqrt{\left(  \frac{E}{g_{2}\left(  E/E_{P}\right)  }\right)
^{2}-m_{0}^{2}\left(  r\right)  }dE. \label{I--}%
\end{equation}
The above integrals can be easily evaluated using the auxiliary variable%
\begin{equation}
z\left(  E/E_{P}\right)  =\frac{E/E_{P}}{g_{2}\left(  E/E_{P}\right)  }%
\end{equation}
so that Eq.$\left(  \ref{LoverG}\right)  $ becomes:
\begin{equation}
\frac{\Lambda}{8\pi G}=-\frac{E_{P}^{4}}{\pi^{2}}\left\{  \int_{x}^{z_{\infty
}}z^{2}\sqrt{z^{2}-x^{2}}dz+\int_{0}^{z_{\infty}}z^{2}\sqrt{z^{2}+x^{2}%
}dz\right\}  , \label{g1g2inv}%
\end{equation}
where $z_{\infty}=\lim_{E\rightarrow\infty}z(E/E_{P})$ and $x=\sqrt{m_{0}%
^{2}\left(  r\right)  /E_{P}^{2}}$. The integrals involved in Eq.$\left(
\ref{g1g2inv}\right)  $ can be calculated straightforwardly being:%
\begin{equation}
I_{1,x}(z)=\int z^{2}\sqrt{z^{2}-x^{2}}dz=\frac{1}{8}\left\{  z\left(
2z^{2}-x^{2}\right)  \sqrt{z^{2}-x^{2}}-x^{4}\log\left[  2\left(
z+\sqrt{z^{2}-x^{2}}\right)  \right]  \right\}  \label{I1m}%
\end{equation}
and%
\begin{equation}
I_{2,x}(z)=\int z^{2}\sqrt{z^{2}+x^{2}}dz=\frac{1}{8}\left\{  z\left(
2z^{2}+x^{2}\right)  \sqrt{z^{2}+x^{2}}-x^{4}\log\left[  2\left(
z+\sqrt{z^{2}+x^{2}}\right)  \right]  \right\}  . \label{I2m}%
\end{equation}

Thus we get the final expression:%
\begin{equation}
\frac{\Lambda}{8\pi G}=-\frac{E_{P}^{4}}{\pi^{2}}\left\{  I_{1,x}(z_{\infty
})-I_{1,x}(x)+I_{2,x}(z_{\infty})-I_{2,x}(0)\right\}  ,
\end{equation}
for the case with $x<z_{\infty}$, and the expression:%
\begin{equation}
\frac{\Lambda}{8\pi G}=-\frac{E_{P}^{4}}{\pi^{2}}\left\{  I_{2,x}(z_{\infty
})-I_{2,x}(0)\right\}  ,
\end{equation}
for the case with $z_{\infty}<x$. In particular for the class of rainbow
functions that satisfy the condition $z_{\infty}=0$ we get a vanishing
cosmological constant:
\begin{equation}
\frac{\Lambda}{8\pi G}=0. \label{Lneg}%
\end{equation}
Although very appealing, the result $\left(  \ref{Lneg}\right)  $ presents the
unpleasant feature of being always negative, even in the region of space where
we would expect a positive cosmological constant. Moreover it is independent
on the choice of $g(E/E_{P})$ provided that this last one can guarantee the
convergence of the integral and the absence of imaginary factors. For these
reasons we are led to investigate other forms of $g_{1}(E/E_{P})$ and
$g_{2}(E/E_{P})$ even if they have less symmetry with respect to proposal
$\left(  \ref{symm}\right)  $. The only restrictions we have are the low
energy limit $\left(  \ref{lim}\right)  $ and the the convergence requirement
for the integrals $\left(  \ref{I+}\right)  $ and $\left(  \ref{I-}\right)  $.
To do calculations in practice a useful choice is the following%
\begin{equation}
g_{1}\left(  E/E_{P}\right)  =\sum_{i=0}^{n}\beta_{i}\frac{E^{i}}{E_{P}^{i}%
}\exp(-\alpha\frac{E^{2}}{E_{P}^{2}}),\qquad g_{2}\left(  E/E_{P}\right)
=1;\qquad\alpha>0,\beta_{i}\in\mathbb{R}. \label{g1g2}%
\end{equation}
The use of a \textquotedblleft Gaussian\textquotedblright\ form is dictated by
the possibility of doing a comparison with NCG models. Indeed, in Ref.\cite{RG
PN} the authors have considered a distortion induced by an underlying NCG on
the counting of states. Basically, one finds that the number of states is
modified in the following way%
\begin{equation}
dn=\frac{d^{3}xd^{3}k}{\left(  2\pi\right)  ^{3}}\ \Longrightarrow
\ dn_{i}=\frac{d^{3}xd^{3}k}{\left(  2\pi\right)  ^{3}}\exp\left(
-\frac{\theta}{4}\left(  \omega_{i,nl}^{2}-m_{i}^{2}\left(  r\right)  \right)
\right)  ,\quad i=1,2, \label{moddn}%
\end{equation}
where the UV cut off is triggered only by higher momenta modes $\gtrsim
1/\sqrt{\theta}$ which propagate over the background geometry. Then the
induced cosmological constant becomes%
\[
\frac{\Lambda}{8\pi G}=\frac{1}{6\pi^{2}}\left[  \int_{\sqrt{m_{0}^{2}\left(
r\right)  }}^{+\infty}\sqrt{\left(  \omega^{2}-m_{0}^{2}\left(  r\right)
\right)  ^{3}}e^{-\frac{\theta}{4}\left(  \omega^{2}-m_{0}^{2}\left(
r\right)  \right)  }d\omega\right.
\]%
\begin{equation}
\left.  +\int_{0}^{+\infty}\sqrt{\left(  \omega^{2}+m_{0}^{2}\left(  r\right)
\right)  ^{3}}e^{-\frac{\theta}{4}\left(  \omega^{2}+m_{0}^{2}\left(
r\right)  \right)  }d\omega\right]  . \label{t1loop}%
\end{equation}
It is immediate to see the analogy with the choice $\left(  \ref{g1g2}\right)
$. However Eq.$\left(  \ref{t1loop}\right)  $ leads directly to a positive
induced cosmological constant, while Eq.$\left(  \ref{LoveG}\right)  $ needs
an appropriate choice of $g_{1}\left(  E/E_{P}\right)  $ and $g_{2}\left(
E/E_{P}\right)  $ to induce a positive part. After choice $\left(
\ref{g1g2}\right)  $, the graviton contribution terms $\left(  \ref{I+}%
\right)  $ and $\left(  \ref{I-}\right)  $ become%
\begin{equation}
I_{+}=3\int_{0}^{\infty}\left(  \sum_{i=0}^{n}\beta_{i}\frac{E^{i}}{E_{P}^{i}%
}\exp(-\alpha\frac{E^{2}}{E_{P}^{2}})\right)  E^{2}\sqrt{E^{2}+m_{0}%
^{2}\left(  r\right)  }dE \label{Iaa+}%
\end{equation}
and%
\begin{equation}
I_{-}=3\int_{\sqrt{m_{0}^{2}\left(  r\right)  }}^{\infty}\left(  \sum
_{i=0}^{n}\beta_{i}\frac{E^{i}}{E_{P}^{i}}\exp(-\alpha\frac{E^{2}}{E_{P}^{2}%
})\right)  E^{2}\sqrt{E^{2}-m_{0}^{2}\left(  r\right)  }dE. \label{Iaa-}%
\end{equation}
In Appendix $\left(  \ref{Integrals}\right)  $, we explicitly compute the
integrals $\left(  \ref{Iaa+}\right)  $, $\left(  \ref{Iaa-}\right)  $ for
every $n$. In order to motivate choice $\left(  \ref{g1g2}\right)  $, we have
to observe that the case with $n=0$ leads to a negative value of $\Lambda/8\pi
G$ for every kind of background as one can see from Eq.$\left(  \ref{LoveG}%
\right)  $. Thus, it is necessary a correction on the pure Gaussian choice in
such a way to have a possible change of sign in $\Lambda/8\pi G$. For our
purposes, it is sufficient to discuss the case with $n=1$ and $n=3$. We begin
with $n=1$.

\subsection{Example a) $\qquad n=1$}

\begin{figure}[h]
\centering\includegraphics[width=2.8in]{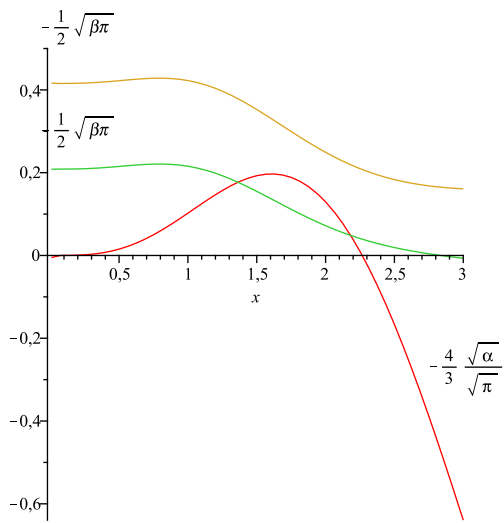}\caption{Plot of
$\Lambda/8\pi G$ as a function of the scale invariant $x$. Choosing
parametrization $\left(  \ref{par1}\right)  $, we obtain the vanishing of
$\Lambda/8\pi G$ when $x\rightarrow0$. The other curves satisfy condition
$\left(  \ref{as}\right)  $ for different values of $\alpha$, with $\alpha
\neq$ $\alpha_{1}$. It is visible the presence of multiple roots.}%
\label{Figs}%
\end{figure}

\begin{figure}[h]
\centering\includegraphics[width=2.8in]{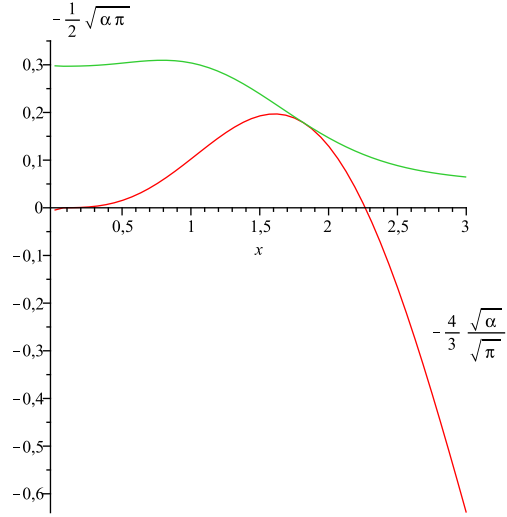}\caption{Plot of
$\Lambda/8\pi G$ as a function of the scale invariant $x$. For $\alpha
_{1}=1/4$, there exists $\alpha_{2}=.7744164292$ where a smooth transition
between the two asymptotic behaviors is possible. The transition appears for
$x=1.818231873$.}%
\label{Fig1}%
\end{figure}

\begin{figure}[h]
\centering\includegraphics[width=3.8in]{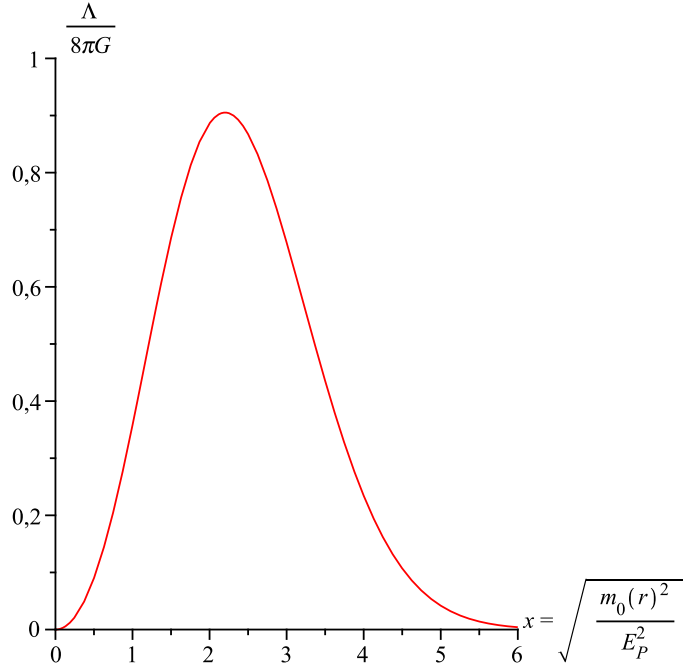}\caption{Plot of
$\Lambda/8\pi G$ as a function of the scale invariant $x$ for $\alpha=1/4.$
The plot works well for backgrounds of the dS, AdS and Minkowski type. Note
that to obtain a positive induced cosmological constant vanishing at small and
large $x$, we need only one parametrization.}%
\label{Fig2}%
\end{figure}

After integration, for $n=1$, Eq.$\left(  \ref{LoveG}\right)  $ can be
rearranged in the following way%
\[
\frac{\Lambda}{8\pi GE_{P}^{4}}\equiv\frac{\Lambda}{8\pi GE_{P}^{4}}\left(
\alpha;\beta;x\right)  =-\frac{1}{2\pi^{2}}\left[  \frac{x^{2}}{\alpha}%
\cosh\left(  \frac{\alpha x^{2}}{2}\right)  K_{1}\left(  \frac{\alpha x^{2}%
}{2}\right)  \right.
\]%
\begin{equation}
\left.  -\beta\left(  {\frac{3x}{2{\alpha}^{2}}}-\frac{x^{2}\sqrt{\pi}%
}{{\alpha}^{\frac{3}{2}}}\sinh\left(  \alpha x^{2}\right)  +\frac{3\sqrt{\pi}%
}{2{\alpha}^{\frac{5}{2}}}\cosh\left(  \alpha x^{2}\right)  +\frac{\sqrt{\pi}%
}{2{\alpha}^{\frac{3}{2}}}\left(  x^{2}-\,{\frac{3}{2{\alpha}}}\right)
\,e{^{\alpha x^{2}}}\operatorname{erf}\left(  \sqrt{\alpha\,}x\right)
\right)  \right]  {,} \label{LG}%
\end{equation}
where, again, $x=\sqrt{m_{0}^{2}\left(  r\right)  /E_{P}^{2}}$ , $\beta
_{1}\equiv\beta$ and where $K_{0}\left(  x\right)  $ is the Bessel function
and $\operatorname{erf}\left(  x\right)  $ is the error function. It is clear
that for every choice of the couple $\left(  \alpha,\beta\right)  $ there
exists a curve with a different behavior. Therefore to fix ideas, we will fix
the Gaussian factor $\alpha$ to the same one proposed by the NCG setting of
Eq.$\left(  \ref{t1loop}\right)  $. Before doing this, it is useful to compute
the series expansion for small and large $x$. For large $x$ one gets%
\begin{equation}
\frac{\Lambda}{8\pi GE_{P}^{4}}\simeq-{\frac{\left(  2\beta{\alpha}%
^{3/2}+\sqrt{\pi}{\alpha}^{2}\right)  x}{4\pi^{2}{\alpha}^{7/2}}-\frac
{8\beta{\alpha}^{5/2}+3\sqrt{\pi}{\alpha}^{3}}{16\pi^{2}{\alpha}^{11/2}%
x}+\frac{3}{128\pi^{2}}}\,{\frac{16\beta{\alpha}^{7/2}+5\sqrt{\pi}{\alpha}%
^{4}}{{\alpha}^{15/2}{x}^{3}}}+O\left(  x^{-4}\right)  , \label{AsL}%
\end{equation}
while for small $x$ we obtain%
\begin{equation}
\frac{\Lambda}{8\pi GE_{P}^{4}}\simeq-{\frac{4{\alpha}^{5/2}+3\sqrt{\pi}%
\beta{\alpha}^{2}}{4\pi^{2}{\alpha}^{9/2}}}+O\left(  x^{3}\right)  .
\label{SmL}%
\end{equation}
It is straightforward to see that if we set%
\begin{equation}
\beta=-{\frac{\sqrt{\alpha\pi}}{2}}, \label{as}%
\end{equation}
then the linear divergent term of the asymptotic expansion $\left(
\ref{AsL}\right)  $ disappears and Eq. $\left(  \ref{LG}\right)  $ vanishes
for large $x$. Plugging Eq.$\left(  \ref{as}\right)  $ into expansion $\left(
\ref{SmL}\right)  $, we obtain%
\begin{equation}
\frac{\Lambda}{8\pi GE_{P}^{4}}\simeq{\frac{{3\pi-8}}{8\pi^{2}{\alpha}^{2}}%
}+O\left(  x^{3}\right)  , \label{AsLSm}%
\end{equation}
which means that for $x=0$, the induced cosmological constant never vanishes
and therefore cannot be a good candidate to reproduce the Minkowskian limit.
Indeed, we have to recall that the variable $x$ expresses the curvature of the
background through the shape function $b\left(  r\right)  $, which for
Minkowski vanishes. On the other hand, the vanishing of expression $\left(
\ref{AsL}\right)  $ and consequently Eq.$\left(  \ref{LG}\right)  $ for
$x\rightarrow\infty$ offers a good candidate for large distances estimates.
Alternatively, by imposing that%
\begin{equation}
\beta=-{\frac{4\,}{3}}\sqrt{\frac{\alpha}{\pi}}, \label{sm}%
\end{equation}
the expression $\left(  \ref{SmL}\right)  $ vanishes for small $x$, while for
large $x$, the leading term becomes%
\begin{equation}
\frac{\Lambda}{8\pi GE_{P}^{4}}\simeq-{\frac{\left(  3\pi-8\right)  x}%
{12\sqrt{\left(  \pi{\alpha}\right)  ^{3}}}.} \label{SmLAs}%
\end{equation}
This means that Eq.$\left(  \ref{LG}\right)  $ diverges towards negative
values. It is straightforward to see that we cannot simultaneously fix both
the conditions $\left(  \ref{as}\right)  $ and $\left(  \ref{sm}\right)  $ for
the same $\alpha$ in order to have a vanishing expectation value of
$\Lambda/8\pi G$ for small and large $x$ unless we consider different values
for $\alpha$ for the different behaviors. The idea is to find a point where a
transition from one parametrization to the other one exists. To begin we have
to observe that if we fix one couple of parameters to%
\begin{equation}
\alpha_{1}=\frac{1}{4},\qquad\beta=-{\frac{4\,}{3}}\sqrt{\frac{\alpha_{1}}%
{\pi}}, \label{par1}%
\end{equation}
where $\alpha_{1}$ has the same value of the numerical factor appearing in
Eq.$\left(  \ref{t1loop}\right)  $ and the second couple with generic values,
one discovers multiple roots where a smooth transition from one
parametrization to the other one can happen. This is illustrated in
Fig.\ref{Figs},where the couple $\left(  \ref{par1}\right)  $ together with some
generic values of the couple satisfying condition $\left(  \ref{as}\right)  $
are shown. It is visible the presence of multiple roots. It is also immediate
to see that there exists one and only one transition point which can be found
by imposing the existence of a tangent point between the curves parametrized
by the values in $\left(  \ref{par1}\right)  $ and the curves parametrized by%
\begin{equation}
\alpha_{2},\qquad\beta=-{\frac{\sqrt{\alpha_{2}\pi}}{2}},
\end{equation}
where $\alpha_{2}$ is to be determined.\textbf{ }We end up with the following
choice%
\begin{equation}
\alpha_{3}=.7744164292,\qquad\beta=-{\frac{\sqrt{\alpha_{3}\pi}}{2},}
\label{par2}%
\end{equation}
where the common point is located in $x=1.818231873$ as shown in
Fig.\ref{Fig1},This choice corresponds to the following setting%
\begin{equation}
\left\{
\begin{array}
[c]{cc}%
g_{1}\left(  E/E_{P}\right)  =\exp(-\alpha_{1}\frac{E^{2}}{E_{P}^{2}})\left(
1-\sqrt{\frac{\alpha_{1}}{\pi}}\frac{4E}{3E_{P}}\right)  ,\qquad g_{2}\left(
E/E_{P}\right)  =1 & 0\leq x\leq1.818231873\\
g_{1}\left(  E/E_{P}\right)  =\exp(-\alpha_{2}\frac{E^{2}}{E_{P}^{2}})\left(
1-\frac{\sqrt{\alpha_{2}\pi}E}{2E_{P}}\right)  ,\qquad g_{2}\left(
E/E_{P}\right)  =1 & x\geq1.818231873
\end{array}
\right.  . \label{match}%
\end{equation}
The setting $\left(  \ref{match}\right)  $ allows the expression $\left(
\ref{LoveG}\right)  $ to have finite values for every kind of background of
the spherically symmetric type. Let us apply our result to the Schwarzschild
background. In terms of the variable $x$, we find that%
\begin{equation}
x=\sqrt{\frac{m_{0}^{2}\left(  r\right)  }{E_{P}^{2}}}=\sqrt{\frac{3MG}%
{r^{3}E_{P}^{2}}}=\left\{
\begin{array}
[c]{cc}%
\frac{3MG}{r^{3}E_{P}^{2}} & r>2MG\\
\frac{3}{8\left(  MG\right)  ^{2}E_{P}^{2}} & r=2MG
\end{array}
\right.  .
\end{equation}
Its behavior is%
\begin{equation}
x\rightarrow\left\{
\begin{array}
[c]{c}%
\infty\qquad\text{\textrm{when}}\qquad M\rightarrow0\qquad
\text{\textrm{for\qquad}}r=2MG\\
0\qquad\text{\textrm{when}}\qquad M\rightarrow0\qquad\text{\textrm{for\qquad}%
}r>2MG
\end{array}
\right.  , \label{small}%
\end{equation}
while%
\begin{equation}
x\rightarrow\left\{
\begin{array}
[c]{c}%
0\qquad\text{\textrm{when}}\qquad M\rightarrow\infty\qquad
\text{\textrm{for\qquad}}r=2MG\\
\infty\qquad\text{\textrm{when}}\qquad M\rightarrow\infty\qquad
\text{\textrm{for\qquad}}r>2MG
\end{array}
\right.  . \label{large}%
\end{equation}
The situation with $M\rightarrow\infty$ describes a wormhole incorporating the
whole universe which is not a physical situation, while for $M\rightarrow0$ we
approach the Minkowski limit which should predict a vanishing induced
cosmological constant. Note that for both setting $\left(  \ref{as}\right)  $
and $\left(  \ref{sm}\right)  $, we find that the whole behavior can be
summarized by the following double limit%
\begin{equation}
\lim_{M\rightarrow0}\lim_{r\rightarrow2MG}\frac{\Lambda\left(  r\right)
}{8\pi G}\neq\lim_{r\rightarrow2MG}\lim_{M\rightarrow0}\frac{\Lambda\left(
r\right)  }{8\pi G}, \label{NonComm}%
\end{equation}
suggesting that a sort of non-commutativity emerges in proximity of the
throat. Therefore, when we adopt the parametrization $\left(  \ref{match}%
\right)  $, the Minkowskian limit is recovered for \textit{every value} of
$M$. Turning now to the case of Eq.$\left(  \ref{LoveGeq}\right)  $, we find
that it is possible to have only one parametrization to obtain the desired
behavior as shown in Fig.\ref{Fig2}

\subsection{Example b)$\qquad n=3.$}

The example we want to analyze corresponds to the case $n=3$. Of course we are
not going to discuss all the possible cases. However $n=3$, represents a fair
compromise of generalization. In the region where relation $\left(
\ref{cond}\right)  $ is valid, the integration of Eq.$\left(  \ref{LoveG}%
\right)  $ gives%
\[
\frac{\Lambda}{8\pi GE_{P}^{4}}=\frac{e^{-x^{2}\alpha}}{16\pi^{2}\alpha^{7/2}%
}\left\{  -\sqrt{\pi}\left(  15\gamma+4x^{2}\alpha^{2}\left(  \beta
+x^{2}\gamma\right)  +6\alpha\left(  \beta+2x^{2}\gamma\right)  \right)
\right.
\]%
\[
-2e^{\frac{x^{2}\alpha}{2}}\left(  1+e^{x^{2}\alpha}\right)  x^{4}\alpha
^{5/2}\delta K_{0}\left(  \frac{x^{2}\alpha}{2}\right)  +e^{2x^{2}\alpha}%
\sqrt{\pi}\left(  2\alpha\left(  -3+2x^{2}\alpha\right)  \beta+\left(
-15-4x^{2}\alpha\left(  -3+x^{2}\alpha\right)  \right)  \gamma\right)
\operatorname{erf}\left(  x\sqrt{\alpha}\right)  +
\]%
\begin{equation}
\left.  +2e^{x^{2}\alpha}x\sqrt{\alpha}\left(  -6\alpha\beta-15\gamma
+2x^{2}\alpha\gamma+2x\alpha K_{1}\left(  \frac{x^{2}\alpha}{2}\right)
\left(  -2(\alpha+2\delta)\cosh\left(  \frac{x^{2}\alpha}{2}\right)
+x^{2}\alpha\delta\sinh\left(  \frac{x^{2}\alpha}{2}\right)  \right)  \right)
\right\}  , \label{Lambda4p}%
\end{equation}

where $K_{\nu}\left(  x\right)  $ $\left(  \nu=0,1\right)  $ are the Bessel
functions\textbf{,} $\delta\equiv\beta_{2}$ and $\gamma\equiv\beta_{3}%
$\textbf{. }Instead, in the region where%
\begin{equation}
m_{0}^{2}\left(  r\right)  =m_{2}^{2}\left(  r\right)  =m_{1}^{2}\left(
r\right)  ,\qquad\forall r\in\left(  r_{1},\infty\right)  ,
\end{equation}
we get%
\begin{equation}
\frac{\Lambda}{8\pi GE_{P}^{4}}=\frac{e^{-x^{2}\alpha}}{8\pi^{2}\alpha^{7/2}%
}\left\{  -\sqrt{\pi}\left(  15\gamma+4x^{2}\alpha^{2}\left(  \beta
+x^{2}\gamma\right)  +6\alpha\left(  \beta+2x^{2}\gamma\right)  \right)
\right. \nonumber
\end{equation}%
\begin{equation}
\left.  +2e^{\frac{x^{2}\alpha}{2}}x^{2}\alpha^{3/2}-x^{2}\alpha\delta
K_{0}\left(  \frac{x^{2}\alpha}{2}\right)  -\left(  4\delta+\alpha\left(
2+x^{2}\delta\right)  \right)  K_{1}\left(  \frac{x^{2}\alpha}{2}\right)
\right\}  .
\end{equation}
The asymptotic expansion of Eq.$\left(  \ref{Lambda4p}\right)  $ in the small
$x$ regime is:
\begin{equation}
\frac{\Lambda}{8\pi GE_{P}^{4}}=-\frac{8\alpha^{3/2}+6\sqrt{\pi}\alpha
\beta+15\sqrt{\pi}\gamma+16\sqrt{\alpha}\delta}{8\pi^{2}\alpha^{7/2}}%
+O(x^{3}), \label{ser3_x_small}%
\end{equation}
whereas the leading contributions to Eq. $(\ref{Lambda4p})$ for large $x$ are:%
\[
\frac{\Lambda}{8\pi GE_{P}^{4}}=-\frac{x\left(  2\sqrt{\pi}\alpha
^{3/2}+4\alpha\beta+8\gamma+3\sqrt{\pi}\sqrt{\alpha}\delta\right)  }{8\pi
^{2}\alpha^{3}}-\frac{6\sqrt{\pi}\alpha^{3/2}+16\alpha\beta+48\gamma
+15\sqrt{\pi}\sqrt{\alpha}\delta}{32\pi^{2}x\alpha^{4}}%
\]%
\begin{equation}
+\frac{3\left(  40\sqrt{\pi}\alpha^{3/2}+128\alpha\beta+512\gamma+105\sqrt
{\pi}\sqrt{\alpha}\delta\right)  }{1024\pi^{2}x^{3}\alpha^{5}}+O(x^{-4}).
\label{ser3_x_large}%
\end{equation}

Again, as in the case of $n=1$, we find that there is in principle a leading
linear divergency in the large $x$ regime. However we can choose the
parameters satisfying the Minkowski limit (i.e. the limit of vanishing
cosmological constant density). This time, differently from the $n=1$ case, we
can ask that the Minkowski limit is satisfied both in the $x\rightarrow0$ and
in the $x\rightarrow\infty$\textbf{ }region, with a unique choice of the
parameters. From Eq.$\left(  \ref{ser3_x_small}\right)  $ and Eq.$\left(
\ref{ser3_x_large}\right)  $ follows that the parameters that satisfy these
requests have to solve the system%
\begin{equation}
\left\{
\begin{array}
[c]{c}%
2\sqrt{\pi}\alpha^{3/2}+4\alpha\beta+8\gamma+3\sqrt{\pi}\sqrt{\alpha}%
\delta=0\\
8\alpha^{3/2}+6\sqrt{\pi}\alpha\beta+15\sqrt{\pi}\gamma+16\sqrt{\alpha}%
\delta=0
\end{array}
\right.  \label{sist_n4}%
\end{equation}

\begin{figure}[h]
\centering\includegraphics[width=3.8in]{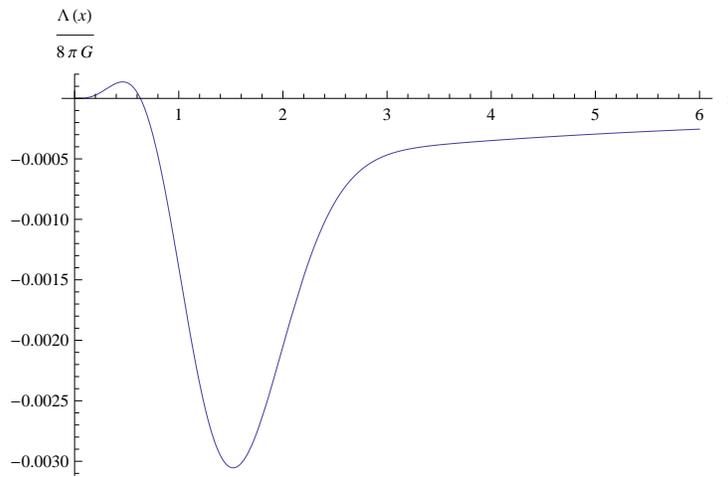}\caption{Plot of
$\Lambda/8\pi G$ as a function of the scale invariant $x$ for $\alpha=1/4,$
$\beta=2\sqrt{\pi}/(9\pi-32)$, $\delta=(8-3\pi)/(18\pi-64)$, $\gamma=0.$ The
plot shows that the Minkowski limit is satisfied both in the $x\rightarrow0$
and in the $x\rightarrow\infty$ limit, but a large $\Lambda<0$ zone is
present.}%
\label{Fig3}%
\end{figure}

Notice that already in the case of three parameters (i.e. $\gamma=0$) the
system $\left(  \ref{sist_n4}\right)  $ can be solved but one gets negative
values of the cosmological constant density in a large $x>1$ zone (see e.g.
Fig.\ref{Fig3}).

Finally, in the full four parameter case (i.e. $\gamma\neq0$), the system
$\left(  \ref{sist_n4}\right)  $ can be solved with the further request of
approaching the Minkowski limit maintaining positive values of $\Lambda.$ A
solution of Eq.$\left(  \ref{sist_n4}\right)  $ satisfying this further
request is:%
\begin{align}
\alpha &  =1/4,\label{n_3}\\
\beta &  =\frac{13635\pi+2048\sqrt{\pi}-38784}{1024(9\pi-32)},\nonumber\\
\delta &  =-\frac{768\pi+909\sqrt{\pi}-2048}{512(9\pi-32)},\nonumber\\
\gamma &  =-\frac{303}{2048}.\nonumber
\end{align}

\begin{figure}[h]
\centering\includegraphics[width=3.8in]{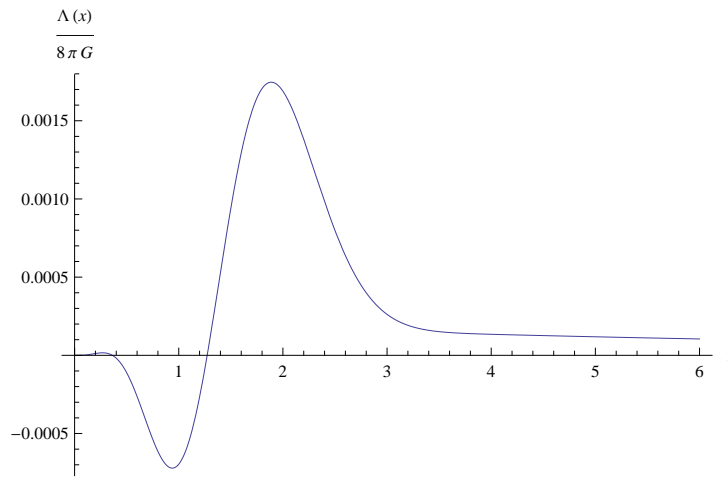}\caption{Plot of
$\Lambda/8\pi G$ as a function of the scale invariant $x$ for values of the
parameters given by Eq.(\ref{n_3}). The plot shows that the Minkowski limit is
satisfied both in the $x\rightarrow0$ and in the $x\rightarrow\infty$ limit.
Both limits are approached maintaining a positive $\Lambda.$ A small
$\Lambda<0$ zone is however still present.}%
\label{Fig4}%
\end{figure}

The resulting $\Lambda$ \ as a function of $x$ is plotted is Fig.\ref{Fig4}. A
small zone in which the cosmological constant density maintains a negative
value is however still present.

\section{Summary and discussion}

\label{p4}Motivated by the promising results obtained in the application of
Gravity's Rainbow to Black Hole Entropy computation\cite{RemoPLB} and, on the
other side in NCG application of ZPE evaluation\cite{RG PN}, in this paper we
have considered how Gravity's Rainbow influences the UV behavior of ZPE. We
have found that, due to the arbitrariness of $g_{1}\left(  E/E_{P}\right)  $
and $g_{2}\left(  E/E_{P}\right)  $, it is always possible to find a form of
the rainbow functions in such a way the expression in $\left(  \ref{VEV}%
\right)  $ be UV finite. As introduced in Ref.\cite{Remo}, the finite result
$\Lambda/8\pi G$ is interpreted as an induced cosmological constant but
without a regularization and a renormalization to keep under control the UV
divergences. To fix ideas we have used Gaussian regulators. In this way our
approach is directly comparable to NCG . The first evident difference is that
in NCG the regulator comes into play into the counting of nodes, while in
Gravity's Rainbow appear in both the sum over eigenvalues and in the counting
of nodes. If one fixes the attention on the pure Gaussian regulator one
discovers that the ZPE is always negative for Gravity's Rainbow. This
unpleasant feature can be corrected with the introduction of a polynomial with
real arbitrary coefficients like in Eq.$\left(  \ref{g1g2}\right)  $. By
imposing the positivity of the result for every $x=\sqrt{m_{0}^{2}\left(
r\right)  /E_{P}^{2}}$ we find that, in case condition $\left(  \ref{equal}%
\right)  $ be satisfied, the parametrization $\left(  \ref{sm}\right)  $ is
sufficient to guarantee that $\Lambda/8\pi G>0$ as shown in Fig.\ref{Fig3}. On
the other hand, when condition $\left(  \ref{cond}\right)  $ is satisfied we
need two different parametrizations to guarantee a correct behavior of
$\Lambda/8\pi G$ for $x\in\left(  0,+\infty\right)  $ and most importantly a
point of connection where a smooth transition can happen as shown in
parametrization $\left(  \ref{match}\right)  $ and in Fig.\ref{Fig2}. In
summary, the final plot becomes\begin{figure}[h]
\centering\includegraphics[width=3.8in]{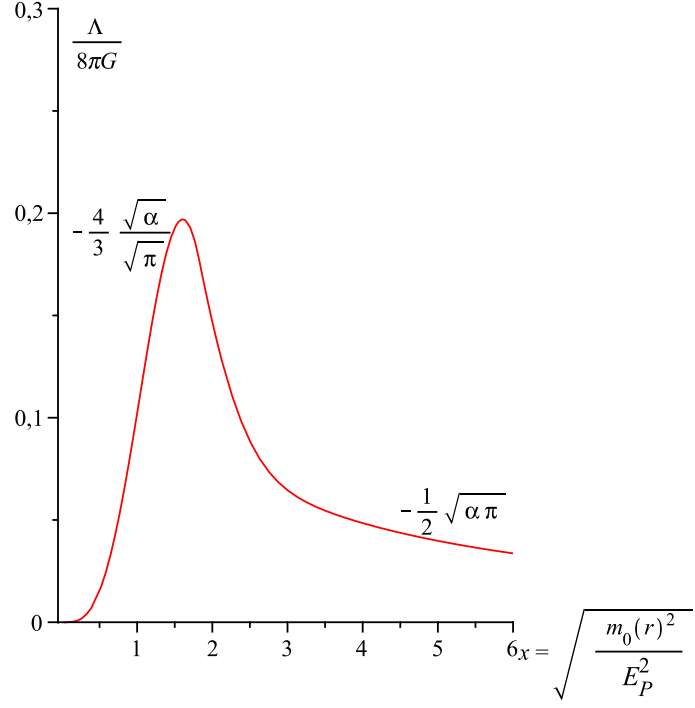}\caption{Plot of
$\Lambda/8\pi G$ as a function of the scale invariant $x$ for values of the
parameters given by Eq.$\left(  \ref{match}\right)  $. The plot shows that the
Minkowski limit is satisfied at the boundaries of the range $\left(
0,+\infty\right)  $. Both limits are approached maintaining a positive
$\Lambda.$ The undesired part of the plot has been eliminated to visualize the
global behavior.}%
\label{fusion}%
\end{figure}Note that the transition clearly highlights that we need two
metrics with the same background $b\left(  r\right)  $ but with two different
choices of $g_{1}\left(  E/E_{P}\right)  $ and $g_{2}\left(  E/E_{P}\right)
$. Of course this transition happens because we insist to have positivity and
a vanishing behavior at the boundary of the range $\left(  0,+\infty\right)
$. The vanishing behavior for small $x$ is a guarantee that the Minkowski
limit is reproduced. Note that the reproduction of a Minkowski limit in NCG
(i.e. the $\sqrt{\theta}\rightarrow0$ limit) is less trivial because of the
existence of the IR/UV mixing\cite{IRUV}. It also appears that the case $n=1$
seems to be special because it is the only one that has a positive
$\Lambda/8\pi G$ for $x\in\left(  0,+\infty\right)  $ when condition $\left(
\ref{cond}\right)  $ is satisfied. We can focus our attention on this case and
suppose to consider the de Sitter background which is the static
representation of the Friedmann-Robertson-Walker model. For this choice, the
shape function $b\left(  r\right)  $ is%
\begin{equation}
b\left(  r\right)  =\frac{\Lambda_{dS}}{3}r^{3}%
\end{equation}
and the effective masses $\left(  \ref{masses}\right)  $ become on the
cosmological throat $r_{c}=\sqrt{3/\Lambda_{dS}}$%
\begin{equation}
m_{0}^{2}\left(  r\right)  =m_{2}^{2}\left(  r\right)  =m_{1}^{2}\left(
r\right)  =\Lambda_{dS}.
\end{equation}
Since the behavior of $\Lambda/8\pi G$ for the de Sitter universe is described
by Fig.\ref{Fig2}, we can compare our results with observation. Since
$\Lambda/8\pi G$ represents the observed cosmological constant induced by
quantum fluctuations of the pure gravitational field, we can fix its value at
the present day as%
\begin{equation}
\frac{\Lambda}{8\pi G}\simeq10^{-11}eV^{4}%
\end{equation}
which can be described either for $x\ll1$ or $x\gg1$. However, for the de
Sitter case $x=\sqrt{\Lambda_{dS}}/E_{P}$ and $x\ll1$ means $r_{c}\gg1$ which
is in agreement with our present universe. On the other hand, when $x\gg1$
means that $r_{c}\ll1$ which could be in agreement with the very early
universe except for the disagreement with the expected theoretical prediction
which for our plot in units of $E_{P}^{4}$ should be $O\left(  1\right)  $.
Therefore it appears that only the left branch of Fig.\ref{Fig2} from the
bottom to the hilltop can be interpreted as a sort of a \textquotedblleft%
\textit{backward evolution}\textquotedblright\ in the radial coordinate $r$.
However to follow the curve from $x\simeq2.18$ to $x\simeq0$ one should have a
variable $\Lambda_{dS}$, namely $\Lambda_{dS}\equiv\Lambda_{dS}\left(
r\right)  $. The same situation appears to exist for Minkowski space in radial
coordinates and for the AdS space. Fortunately, Minkowski space has no a
preferred scale and Fig.\ref{Fig2} has the correct asymptotic behavior except
for an unpleasant peak in correspondence of the peak location of the dS space.
It is likely that this spurious prediction is due to the coordinate choice. On
the other side one can verify that $\Lambda/8\pi G\rightarrow0$ when
$r\rightarrow0$ for both Minkowski, dS and AdS spaces. Coming back on the AdS
space, we have to note that this background is not endowed of a horizon and
therefore it makes difficult to find a significant point. Nevertheless looking
once again Fig.\ref{Fig2}, we can claim that for $r\rightarrow\infty$ and very
small $\Lambda_{AdS}$ one gets%
\begin{equation}
b\left(  r\right)  =-\frac{\Lambda_{AdS}}{3}r^{3}\qquad\mathrm{and\qquad
}x=\sqrt{\frac{6/r^{2}+\Lambda_{AdS}}{E_{P}^{2}}}\rightarrow0,
\end{equation}
namely a vanishing $\Lambda/8\pi G$, while in the other regime, i.e.
$r\rightarrow0$, $x\rightarrow\infty$ and once again one obtains a vanishing
$\Lambda/8\pi G$ that it means that against all odds, we have regularity on
the singularity $r=0$.

\appendix{}

\section{Integrals}

\label{Integrals}In this appendix, we explicitly compute the integrals coming
from Eq.$\left(  \ref{LoveG}\right)  $. We begin with%
\begin{equation}
I_{+}=3\int_{0}^{\infty}\left[  \sum_{i=0}^{n}c_{i}\frac{E^{i}}{E_{P}^{i}}%
\exp(-\alpha\frac{E^{2}}{E_{P}^{2}})\right]  E^{2}\sqrt{E^{2}+m_{0}^{2}\left(
r\right)  }dE.
\end{equation}
It is useful to divide $I_{+}$ into two pieces with $i$ odd and $i$ even, thus
we can write%
\begin{equation}
I_{+}=I_{+}^{e}+I_{+}^{o},
\end{equation}

where
\begin{align}
I_{+}^{e}  &  =3/2E_{P}^{4}\sum_{i=0}^{n}c_{i}\left(  -\right)  ^{i}%
\lim_{\beta\rightarrow0}\frac{d^{i}}{d\beta^{i}}\int_{0}^{\infty}x^{1/2}%
\exp\left[  -(\alpha+\beta)x\right]  \sqrt{x+m_{0}^{2}\left(  r\right)
/E_{P}^{2}}dx,\\
I_{+}^{o}  &  =3/2E_{P}^{4}\sum_{i=0}^{n}c_{i}\left(  -\right)  ^{i}%
\lim_{\beta\rightarrow0}\frac{d^{i}}{d\beta^{i}}\int_{0}^{\infty}x\exp\left[
-(\alpha+\beta)x\right]  \sqrt{x+m_{0}^{2}\left(  r\right)  /E_{P}^{2}}dx,
\end{align}
are expressed in terms of the variable $x=E^{2}/E_{P}^{2}.$

The integrals involved in the expressions of $I_{+}^{e}$ and $I_{+}^{o}$ \ can
be evaluated using the formulas%
\begin{align}
\int_{0}^{\infty}dx\left(  x+t\right)  ^{1/2}x^{1/2}\exp\left(  -\mu x\right)
&  =\frac{t}{2\mu}\exp\left(  \frac{t\mu}{2}\right)  K_{1}\left(  \frac{t\mu
}{2}\right)  \qquad t>0,\mu>0\\
\int_{0}^{\infty}dx\left(  x+t\right)  ^{1/2}x\exp\left(  -\mu x\right)   &
=\frac{3}{2}\frac{\sqrt{t}}{\mu^{2}}+\frac{\sqrt{\pi}}{4}\mu^{-5/2}\exp
(t\mu)(3-2t\mu)\text{Erfc}\left[  \sqrt{t\mu}\right]  \qquad t>0,\mu>0
\end{align}
obtaining%
\begin{align}
I_{+}^{e}  &  =3/2E_{P}^{4}\sum_{i=0}^{n}c_{i}\left(  -\right)  ^{i}%
\lim_{\beta\rightarrow0}\frac{d^{i}}{d\beta^{i}}\left\{  \frac{m_{0}%
^{2}\left(  r\right)  }{2E_{P}^{2}(\alpha+\beta)}\exp\left[  \frac{m_{0}%
^{2}\left(  r\right)  (\alpha+\beta)}{2E_{P}^{2}}\right]  K_{1}\left[
\frac{m_{0}^{2}\left(  r\right)  }{2E_{P}^{2}}(\alpha+\beta)\right]  \right\}
,\\
I_{+}^{o}  &  =+3/2E_{P}^{4}\sum_{i=0}^{n}c_{i}\left(  -\right)  ^{i}%
\lim_{\beta\rightarrow0}\frac{d^{i}}{d\beta^{i}}\left\{  \frac{3m_{0}\left(
r\right)  }{2E_{P}(\alpha+\beta)^{2}}+\right. \nonumber\\
&  \left.  -\frac{\sqrt{\pi}}{4}\exp\left[  \frac{m_{0}^{2}\left(  r\right)
(\alpha+\beta)}{E_{P}^{2}}\right]  \left[  \frac{2m_{0}^{2}\left(  r\right)
}{E_{P}^{2}(\alpha+\beta)^{3/2}}-\frac{3}{(\alpha+\beta)^{5/2}}\right]
\text{Erfc}\left[  \frac{m_{0}\left(  r\right)  }{E_{P}}\sqrt{(\alpha+\beta
)}\right]  \right\}  .
\end{align}

The same procedure \ can be followed to evaluate%
\begin{equation}
I_{-}=3\int_{\sqrt{m_{0}^{2}\left(  r\right)  }}^{\infty}\left[  \sum
_{i=0}^{n}c_{i}\frac{E^{i}}{E_{P}^{i}}\exp(-\alpha\frac{E^{2}}{E_{P}^{2}%
})\right]  E^{2}\sqrt{E^{2}-m_{0}^{2}\left(  r\right)  }dE.
\end{equation}

Even in this case, it is useful to divide $I_{-}$ into two pieces with $i$ odd
and $i$ even. Thus, we can write%
\begin{equation}
I_{-}=I_{-}^{o}+I_{-}^{e},
\end{equation}
where $I_{-}^{e}$ and $I_{-}^{o}$ are given by%
\begin{align}
I_{-}^{e}  &  =3/2E_{P}^{4}\sum_{i=0}^{n}c_{i}\left(  -\right)  ^{i}%
\lim_{\beta\rightarrow0}\frac{d^{i}}{d\beta^{i}}\int_{m_{0}^{2}/E_{P}^{2}%
}^{\infty}x^{1/2}\exp\left[  -(\alpha+\beta)x\right]  \sqrt{x-m_{0}^{2}\left(
r\right)  /E_{P}^{2}}dx,\\
I_{-}^{o}  &  =3/2E_{P}^{4}\sum_{i=0}^{n}c_{i}\left(  -\right)  ^{i}%
\lim_{\beta\rightarrow0}\frac{d^{i}}{d\beta^{i}}\int_{m_{0}^{2}/E_{P}^{2}%
}^{\infty}x^{1/2}\exp\left[  -(\alpha+\beta)x\right]  \sqrt{x-m_{0}^{2}\left(
r\right)  /E_{P}^{2}}dx.
\end{align}
Using now the formulas%
\begin{align}
\int_{t}^{\infty}dx\left(  x-t\right)  ^{1/2}x^{1/2}\exp\left(  -\mu x\right)
&  =\frac{t}{2\mu}\exp\left(  -\frac{t\mu}{2}\right)  K_{1}\left(  \frac{t\mu
}{2}\right)  \qquad t>0,\mu>0\\
\int_{t}^{\infty}dx\left(  x-t\right)  ^{1/2}x\exp\left(  -\mu x\right)   &
=\frac{\sqrt{\pi}}{4}\mu^{-5/2}(3+2\mu t)\exp\left(  -\mu t\right)  \qquad
t>0,\mu>0
\end{align}
\bigskip$I_{-}^{e}$ and $I_{-}^{o}$ can be rewritten in the form%
\begin{align}
I_{-}^{e}  &  =3/2E_{P}^{4}\sum_{i=0}^{n}c_{i}\left(  -\right)  ^{i}%
\lim_{\beta\rightarrow0}\frac{d^{i}}{d\beta^{i}}\left\{  \frac{m_{0}%
^{2}\left(  r\right)  }{2(\alpha+\beta)E_{P}^{2}}\exp\left[  -\frac{m_{0}%
^{2}\left(  r\right)  (\alpha+\beta)}{2E_{P}^{2}}\right]  K_{1}\left[
\frac{m_{0}^{2}\left(  r\right)  (\alpha+\beta)}{2E_{P}^{2}}\right]  \right\}
,\\
I_{-}^{o}  &  =3/2E_{P}^{4}\sum_{i=0}^{n}c_{i}\left(  -\right)  ^{i}%
\lim_{\beta\rightarrow0}\frac{d^{i}}{d\beta^{i}}\left\{  \frac{\sqrt{\pi}}%
{4}(\alpha+\beta)^{-5/2}\left[  3+2(\alpha+\beta)\frac{m_{0}^{2}\left(
r\right)  }{E_{P}^{2}}\right]  \exp\left[  -\frac{m_{0}^{2}\left(  r\right)
(\alpha+\beta)}{E_{P}^{2}}\right]  \right\}  .
\end{align}


\begin{thebibliography}{99}                                                                                               %


\bibitem {GAC}G. Amelino-Camelia, \textsl{Int.J.Mod.Phys. }\textbf{D 11}, 35
(2002); gr-qc/0012051. G. Amelino-Camelia, \textsl{Phys.Lett. }\textbf{B 510},
255 (2001); hep-th/0012238.

\bibitem {MagSmo}J. Magueijo and L. Smolin, \textsl{Class. Quant. Grav.}
\textbf{21}, 1725 (2004) [arXiv:gr-qc/0305055].

\bibitem {RemoPLB}R. Garattini, Phys.Lett. \textbf{B} 685 329 (2010);
arXiv:0902.3927 [gr-qc].

\bibitem {DeWitt}B. S. DeWitt, \textsl{Phys. Rev.} \textbf{160}, 1113 (1967).

\bibitem {Remo}R. Garattini, \textsl{TSPU Vestnik} \textbf{44} \textbf{N7}, 72
(2004); [arXiv:gr-qc/0409016]. R.Garattini, \textsl{J. Phys. A }\textbf{39},
6393 (2006); [arXiv:gr-qc/0510061]. R. Garattini, \textsl{J.Phys.Conf.Ser}.
\textbf{33}, 215 (2006); [arXiv:gr-qc/0510062].

\bibitem {CG}S. Capozziello and R. Garattini, \textsl{Class.Quant.Grav.}
\textbf{24}, 1627 (2007); gr-qc/0702075.

\bibitem {Vassilevich}D.V. Vassilevich, \textsl{Int. J. Mod. Phys}. \textbf{A
8}, 1637 (1993).

\bibitem {Quad}M. Berger and D. Ebin, \textsl{J.Diff.Geom.} \textbf{3}, 379
(1969). J. W. York Jr., \textsl{J.Math.Phys.}, \textbf{14}, 4 (1973);
\textsl{Ann. Inst. Henri Poincar\'{e}} \textbf{A} \textbf{21}, 319 (1974).

\bibitem {Remo1}R. Garattini, \textit{The Cosmological constant and the
Wheeler-DeWitt Equation}. arXiv:0910.1735 [gr-qc]

\bibitem {Regge Wheeler}T. Regge and J. A. Wheeler, \textsl{Phys.Rev.
}\textbf{108}, 1063 (1957).

\bibitem {tHooft}G. 't Hooft, Nucl. Phys. \textbf{B} \textbf{256}, 727 (1985).

\bibitem {g1g2}Y. Ling, \textsl{JCAP} \textbf{17}, 708 (2007), gr-qc/0609129;
Y. Ling, X. Li and H. Zhang, \textsl{Mod.Phys.Lett. }\textbf{A 22}, 2749
(2007), gr-qc/0512084; Y. Ling, B. Hu and X. Li, \textsl{Phys. Rev.} \textbf{D
73}, 087702 (2006), gr-qc/0512083.

\bibitem {RG PN}R. Garattini and P. Nicolini, \textit{A noncommutative
approach to the cosmological constant problem. }arXiv:1006.5418 [gr-qc].

\bibitem {IRUV}S. Minwalla, M.Van Raamsdonk and N. Seiberg, \textsl{JHEP}
\textbf{020}, 02 (2000), hep-th/9912072; G.Amelino-Camelia, G. Mandanici and
K. Yoshida, \textsl{JHEP} \textbf{037,} 0401 (2004), hep-th/0209254.
\end{thebibliography}
\end{document}